\documentclass[aps,prb,floatfix,twocolumn,showpacs]{revtex4}
\usepackage{epsfig,amsmath,amssymb,rotating,multirow}

\begin{document}

\hsize\textwidth\columnwidth\hsize\csname@twocolumnfalse\endcsname

\title{Control of electron spin and orbital resonance in quantum dots through spin-orbit interactions.}

\author{Peter Stano$^{1,2}$ and Jaroslav Fabian$^1$}
\affiliation{$^1$Institute for Theoretical 
Physics, University of Regensburg, 93040 Regensburg, Germany\\
$^2$Research Center for Quantum Information, Slovak Academy of Sciences, Bratislava, Slovakia}

\vskip1.5truecm
\begin{abstract}
Influence of resonant oscillating electromagnetic field on a single electron in coupled lateral quantum dots in the presence of phonon-induced relaxation and decoherence is investigated. Using symmetry arguments it is shown that spin and orbital resonance can be efficiently controlled by spin-orbit interactions. The control is possible due to the strong sensitivity of Rabi frequency to the dot configuration (orientation of the dot and a static magnetic field) as a result of the anisotropy of the spin-orbit interactions. The so called easy passage configuration is shown to be particularly suitable for magnetic manipulation of spin qubits, ensuring long spin relaxation time and protecting the spin qubit from electric field disturbances accompanying on-chip manipulations.

\end{abstract}
\maketitle

\section{Introduction}

Spin properties of few electron quantum dots have been recently extensively studied, in hope that a localized spin can serve as a qubit, a central building block of a quantum computer.\cite{loss1998:PRA,cerletti2005:N} Spin, compared to orbital degrees of freedom, was anticipated to have much longer coherence time. Fast experimental progress during last few years supported this assumption -- long electron spin relaxation\cite{elzerman2004:N,amasha2006:CM,meunier2007:PRL} and dephasing times\cite{petta2005:S,laird2006:PRL} have been measured in quantum dots.

If a quantum dot electron spin is to realize a qubit, DiVincenzo's criteria have to be fulfilled:\cite{DiVincenzo2000:FP} (i) The existence of a {\it qubit} -- the two states of spin naturally encode the information bit. (ii) The {\it initialization} of the qubit is also straightforward -- at a finite static magnetic field and small temperature it is enough to let the system relax into the ground state spontaneously. (iii) Due to the isolation of the spin from the environment, the qubit {\it readout} is not that easy, but can be now considered experimentally mastered, using spin-to-charge conversion schemes.\cite{engel2004:PRL,hanson2005:PRL} (iv) Concerning coherent {\it manipulation}, a very important step forward is a recent demonstration of magnetically driven Rabi oscillations.\cite{koppens2006:N} Thus, all basic ingredients have been shown to work at the proof of principle level, and the effort now is aimed at their integration, with the final goal of a (v) {\it scalable} qubit design. Connected with the last two points, namely, if dots in an array can be addressed individually, the manipulating fields have to be produced locally, nearby the particular dot being manipulated (so called fields generated on-chip).

If the spin is manipulated by a magnetic field which is produced locally (by an oscillating current in a wire nearby the dot), it is inevitably accompanied by an oscillating electric field. This electric field is due to an imperfect screening of the dot from the circuitry; the electric field due to a changing magnetic field, ${\bf \nabla}\times {\bf E}=-\partial_t {\bf B}$ is negligible.\cite{golovach2006:PRB} The electric field strongly disturbs the orbital part of the electron wavefunction and, if spin-orbit coupling is present, also couples to the spin -- one way or the other, it limits the strength of the magnetic field usable for the manipulation (in Ref.~\onlinecite{koppens2006:N} this limit was 1.9 mT) and thus limits speed of the operation (the maximal achievable Rabi frequency). 

On the other hand, the electric field does not have to be viewed as an enemy -- through the spin-orbit interaction it induces the very same spin oscillations\cite{rashba2003:PRL,golovach2006:PRB} as the magnetic field. From practical point of view, the electric field is preferred, since it is much easier to control than the magnetic field. The possibility of electrically induced spin oscillations is eagerly pursued experimentally. The disadvantage of the electric field is that the Rabi frequency is expected to depend on dot parameters, since the coupling is due to material dependent spin-orbit interactions. (In the case of the magnetic field, the frequency is given only by the field strength.)

It is thus an important issue to compare the effectiveness of oscillating electric and magnetic fields in inducing Rabi spin oscillations. Namely, how large fields are required to induce Rabi oscillation with certain frequency and how the frequency depends on the parameters of the dot (consequently, how stable it is against fluctuations of these parameters). This is where this paper aims -- we quantify the dipolar electric and magnetic couplings in spin resonance of a single electron confined in a quantum dot. By spin resonance we mean that a static magnetic field is applied, whereby the ground orbital state is split by the Zeeman energy. Oscillations between the two split states are induced by oscillating magnetic and electric fields if the field frequency equals to the Zeeman energy.

It was already showed theoretically in single dots that due to the presence of spin-orbit interactions the electric field is indeed effective in inducing spin resonance.\cite{golovach2006:PRB} For a typical lateral single GaAs quantum dot, in static magnetic field of 1 Tesla, an oscillating electric field of $10^3$ V/m is as effective as the oscillating magnetic field of 1 mT. We widen the analysis on the experimentally relevant double dot setup. Our main result is that the anisotropy of the spin-orbit interactions allows for a strong control over the electrical field efficiency in spin manipulations. Our findings provide guide for dot configurations for two possible strategies: If a local electric field is chosen for the spin manipulation, we show how its efficiency can be maximized. On the other hand, if a magnetic field is chosen, the coupling due to the accompanying electric field is unwanted -- we show that it can be suppressed by: (i) lowering the magnitude of the static magnetic field, (ii) special orientation of the static magnetic field. Particularly, in an easy passage configuration,\cite{stano2006:PRL} the otherwise most effective electric field component is completely blocked in disturbing the spin. 

In addition to spin resonance, we also study the electrically and magnetically induced orbital resonance, where the resonant states are the two lowest orbital states with the same spin. A qubit represented by these two states is called a charge qubit. The study is motivated by an observation, that in the presence of spin-orbit interactions, an analogue to electrically induced spin resonance should exist. Namely, the magnetic field should induce oscillation between spin alike states. We show that this is indeed true, however, for realistic values the magnetic field is much less effective compared to the electric field. 

We use realistic parameters for electrically defined single and coupled dots in [001] grown GaAs heterostructure. We treat the problem numerically by exact diagonalization of the full electron Hamiltonian. However, for all results we provide explanatory analytical arguments based on an effective spin-orbit Hamiltonian and the degenerate perturbation theory. Our model incorporates the electron relaxation and decoherence rates caused by acoustic phonons in a realistic way; the rates we use have been found to be in a very good agreement with the experimental data for magnetic fields above 1 Tesla both in single\cite{amasha2006:CM,stano2006:PRL} and double dots.\cite{meunier2007:PRL}

The paper is organized as follows: In Sec.~II we describe the model of the electron in the dissipative phonon environment under oscillating electric and magnetic fields. In Sec.~III we derive an effective spin-orbit Hamiltonian which allows symmetry analysis of the problem. With this Hamiltonian we evaluate the matrix elements of oscillating magnetic and electric fields for the case of spin (Sec.~IV) and orbital (Sec.~V) resonance. In Sec.~VI we describe the system in the steady state, where we show how to obtain the Rabi frequency and decoherence from a steady state measurement.

\section{Model}

\begin{figure}
\centerline{\psfig{file=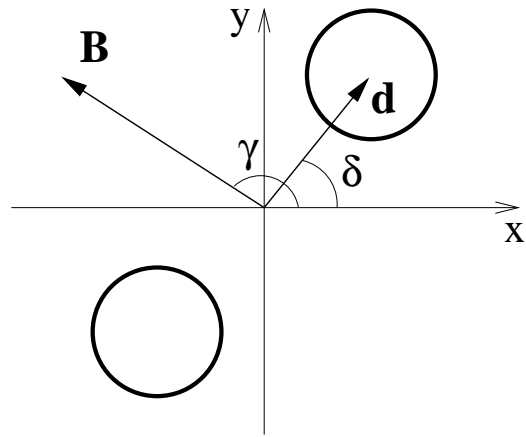,width=0.8\linewidth}}
\caption{The orientation of the potential dot minima (denoted as the two circles) with respect to the crystallographic axes ($x=[100]$ and $y=[010]$) is defined by the angle $\delta$. The orientation of the inplane magnetic field is given by the angle $\gamma$.}
\label{fig:definition}
\end{figure}

Consider a single electron in a double quantum dot formed in a two dimensional electron gas in a (001) plane of a GaAs/AlGaAs heterostructure. The effective Hamiltonian is
\begin{equation} \label{eq:hamiltonian}
H=H_0+H_{BR}+H_{D}+H_{D3},
\end{equation}
where
\begin{equation}\label{eq:h0}
H_0=T+V_{C}+H_Z. 
\end{equation}
The kinetic energy is $T=\hbar \mathbf{k}^2/2m$ with the effective
electron mass $m$ and kinetic momentum $\hbar \mathbf{k} =
-i\hbar\boldsymbol{\nabla}$. 
The double quantum dot is described here by the confinement 
\begin{equation} \label{eq:doubledot} 
V_{C}(\mathbf{r}) = \frac{1}{2} m \omega_0^2
\mathrm{min}\{(\mathbf{r}-\mathbf{d})^2,(\mathbf{r}+\mathbf{d})^2\},
\end{equation}
representing two alike potential minima of a parabolic shape, centered at $\pm \bf{d}$. The in-plane orientation of the double dot with respect to the crystallographic axes $x$ and $y$ is defined by $\delta$, the angle between $\mathbf{d}$ and $\hat{x}$. A single dot with the confinement energy $E_0 = \hbar\omega_0$, and the confinement length $l_0=\surd \hbar/m\omega_0$, is defined by the limit $d=0$. Alternatively to giving the interdot distance $d$, the double dot can be characterized by tunneling energy $\delta E_t$ equal to half of the difference of the energies of the two lowest orbital states.\cite{stano2005:PRB} The electron feels an in-plane magnetic field ${\bf B}$ whose orbital effects can be neglected, for fields lower than $\sim 10$ T. The Zeeman term is $H_Z=\mu\boldsymbol{\sigma}.\mathbf{B}$, where $\mu=(g/2)\mu_B$ is the renormalized magneton, $g$ is the conduction band g-factor, $\mu_B$ is the Bohr magneton, 
and $\boldsymbol{\sigma}$ are the Pauli matrices. The spin quantization axis is defined by the direction of the magnetic field. The angle between ${\bf B}$ and $\hat{x}$ is denoted as $\gamma$.
The geometry is summarized in Fig.~\ref{fig:definition}.

The spin-orbit coupling in our confined
system is described by three terms.\cite{zutic2004:RMP}
The Bychkov-Rashba Hamiltonian,\cite{rashba1960:FTT, bychkov1984:JPC}
\begin{equation}
H_{BR}=\frac{\hbar^2}{2 m l_{BR}}\left(\sigma_x k_y-\sigma_y k_x \right),
\end{equation}
is present due to the heterostructure asymmetry, while the linear and cubic Dresselhaus Hamiltonians,\cite{dresselhaus1955:PR,dyakonov1986:FTP}
\begin{eqnarray}
H_{D}&=&\frac{\hbar^2}{2 m l_D}\left(-\sigma_x k_x+\sigma_y k_y\right),\\
H_{D3}&=&\gamma_c\left(\sigma_x k_x k_y^2-\sigma_y k_y
k_x^2\right),
\end{eqnarray}
are due to the lack of the bulk inversion symmetry.

In our numerical calculations we use bulk GaAs material parameters: 
$m=0.067\,m_e$, $g =-0.44$, and $\gamma_c=27.5$ eV$\mathrm{\AA^3}$. For
the coupling of the linear spin-orbit terms we choose $l_{BR}=1.8$ $\mu$m, and $l_D=1.3$ $\mu$m, the values used to fit a recent experiment.\cite{stano2006:PRL} For the confinement length we take $l_0=30$ nm, corresponding to the confinement energy $E_0=1.2$ meV.

We now describe the influence of the phonon environment as well as of the oscillating electric and magnetic fields. The phonon environment leads to the relaxation and decoherence expressed, in the Markov and Born approximations, by the time derivative of the diagonal and off-diagonal elements of the reduced density matrix of the electron, $\rho$:\cite{blum} (upperscript ``$\rm ph$'' stands for phonons, to discriminate from other contributions to the time derivative which appear later)
\begin{subequations}
\label{eq:phonon channel}
\begin{eqnarray}
\partial_t^{\rm ph}\, \rho_{ii}&=&-\sum_{k} 2 \Gamma_{ik}\rho_{ii}+\sum_{k} 2 \Gamma_{ki}\rho_{kk},
\label{eq:phonon channel a}\\
\partial_t^{\rm ph}\, \rho_{ij}&=&-\sum_{k} (\Gamma_{ik}+\Gamma_{jk}) \rho_{ij}\equiv -\gamma_{ij} \rho_{ij}.
\label{eq:phonon channel b}
\end{eqnarray}  
\end{subequations}

Here $2\Gamma_{ij}$ is the relaxation rate from the electron state $i$ to $j$ 
due to the piezoelectric and deformation potential interactions of the electron with acoustic phonons. There is no additional phonon channel for the decoherence $\gamma_{ij}$ apart from the relaxation, since the phonon density of states vanishes for zero phonon energy, $\Gamma_{ii}= 0$. We do not consider non-phonon mechanisms of dephasing, which are important at low (sub Tesla) magnetic fields. To allow for a finite temperature one can suppose a detailed balance: $\Gamma_{ji}=\tau \Gamma_{ij}$, where $\tau=\exp(-\hbar \omega_{ij}/k_B T)$. In the calculations below, we consider temperature much lower than the orbital excitation energy. For example, the experiment  Ref.~\onlinecite{koppens2006:N} was done at temperature 100 mK, corresponding to $\sim 0.01$ meV, while a typical excitation energy of the used quantum dot was about 1 meV. In this limit a transition into a higher orbital level has a negligible rate. 

In addition to phonons, the electron is subject to oscillating electric and magnetic fields, which contribute through the following Hamiltonian:
\begin{equation} \begin{split}
H^{\rm of}&=[e \boldsymbol{\mathcal{E}}.{\bf r}+\mu \boldsymbol{\mathcal{B}}.\boldsymbol{\sigma}]  \cos \omega t\equiv \hbar \hat{\Omega} \cos \omega t.
\end{split} 
\end{equation}
Only the in-plane components of the oscillating electric field are relevant. The oscillating magnetic field is perpendicular to the plane, $\boldsymbol{\mathcal{B}}\propto \hat{z}$, simulating the conditions in the experiment.\cite{koppens2006:N} In the numerical calculations we set $E=1000$ V/m as a realistic guess for the experimental setup\cite{koppens:P} and $B=1$ mT, a typical value from the experiment.\cite{koppens2006:N} We suppose that frequency $\omega$ is close to the energy difference of a given pair of states -- resonant states -- denoted by indexes $a$ and $b$, such that $\omega\approx \omega_{ba}=(E_b-E_a)/\hbar>0$. In the rotating wave approximation,\cite{blum} that we adopt, the oscillating field influences only the two resonant states, contributing to the time derivative of the density matrix: (upperscript ``of'' stands for the oscillating field)
\begin{subequations}
\label{eq:field channel}
\begin{eqnarray}
\partial_t^{\rm of} \rho_{aa}&=&- \partial_t^{\rm of}\rho_{bb}=
\frac{\rm{i}}{2} \Omega_{ba}\rho_{ab}e^{i\Delta t}-\frac{\rm{i}}{2} \Omega_{ab}\rho_{ba}e^{-\rm{i} \Delta t},
\label{eq:field channel a}\\
\partial_t^{\rm of} \rho_{ab}&=&-\frac{\rm{i}}{2} (\rho_{bb}-\rho_{aa})\Omega_{ab}e^{-\rm{i} \Delta t},
\label{eq:field channel b}
\end{eqnarray}
\end{subequations}
where $\Delta=\omega_{ba}-\omega$ is the detuning from the resonance.

The time evolution of the density matrix, given by Eqs.~\eqref{eq:phonon channel} and \eqref{eq:field channel} can be easily solved if one neglects all other states but the two resonant.\cite{fabian2007:APS} Such approximation makes sense if the electron can not escape from the two state subspace. Roughly speaking, the effective rate out of the subspace must be much smaller than rates for transitions restoring the electron in. This, for example, means that the ground state must be one of the resonant states, which is the case here. Another interesting counterexample is optical shelving,\cite{nagourney1986:PRL} whereby the electron can be trapped in an intermediate dark state. There are parameter values for our model where the three lowest electron states can realize such a scheme, but we do not discuss it in this article. We work in the regime where the two level approximation is justified, as follows also from our numerical results. The validity of the two level approximation also implies that the decoherence rate is given by the relaxation only, 
\begin{equation}
\gamma_{ab}=\gamma_{ba}=\Gamma_{ba}+\Gamma_{ab}=\Gamma_{ba}(1+\tau), 
\label{eq:decoherence} 
\end{equation}
a fact that we will use later. 

Suppose now the electron is in the ground state initially. After the resonant field is turned on, the populations of the two resonant states start to oscillate, meaning, after a certain time the electron will be in the excited state, then comes back to the ground state and so on. Since these Rabi oscillations are coherent, they can realize a single qubit rotation, one of the basic building blocks of a quantum computation. The time after which the populations switch is proportional to the inverse of the frequency of the Rabi oscillations (Rabi frequency) $\Omega$. A larger Rabi frequency then means a faster single qubit operation. To better assess the suitability for quantum computation, one has to take into account also the decay of the Rabi oscillations which is due to the decoherence. In our model, the magnitude of the oscillations decays exponentially with the rate roughly proportional to the decoherence rate $\gamma_{ba}$. Therefore, to minimize the error in a single qubit operation, it is desirable to maximize the ratio $\Omega$/$\gamma_{ba}$, which quantifies how many single qubit operations one can do during the decoherence time. We note that from the observed decaying Rabi oscillations in the time domain,\cite{petta2005:S,koppens2006:N} both $\Omega$ and $\gamma_{ba}$ can be extracted.

As the last here we note that in the two resonant states approximation there are three important rates, decoherence $\gamma_{ba}$, detuning $\Delta$ and the field matrix element $|\Omega_{ab}|$. If the the last one is not dominant, then either $\gamma_{ab}$ is large and the damping is too strong to observe Rabi oscillations or $\Delta$ is large and the magnitude of the oscillations is small\cite{fabian2007:APS} -- both cases are not of interest here. We consider the case when the field matrix element is indeed dominant. It holds then that the matrix element equals the Rabi frequency,
\begin{equation}
\Omega=|\Omega_{ba}|,
\end{equation} 
and is therefore of crucial importance. In the next we analyze in detail how the field matrix element due to electric and magnetic oscillating fields depends on system parameters. To simplify the analysis of the spin-orbit influence, we begin with a derivation of an effective spin-orbit Hamiltonian.

\section{Effective spin-orbit Hamiltonian}

\begin{figure}
\centerline{\psfig{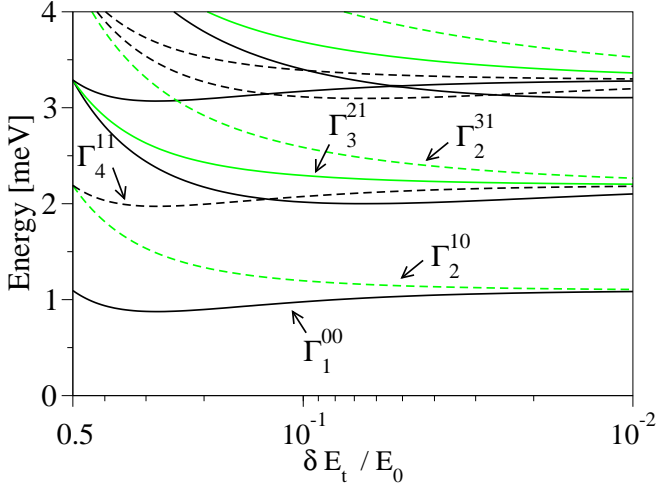}}
\caption{(Color online) The lowest part of the energy spectrum of Hamiltonian $H_0$, Eq.~\eqref{eq:h0}, in zero magnetic field as a function of the tunneling energy ($\delta E_t$) in the units of the confinement energy $E_0$. Each eigenfunction belongs to one of the four symmetry classes of $C_{2v}$, which are denoted by different color/type of the line. Spin indexes are omitted.}
\label{fig:symmetry}
\end{figure}

It is useful to remove the linear spin-orbit terms in Eq.~\eqref{eq:hamiltonian} by applying a unitary transformation,\cite{aleiner2001:PRL} leading to a new Hamiltonian 
\begin{equation}
H^\prime=e^{S} H e^{-S}=H_0+H_1,
\label{eq:unitary transformation}
\end{equation}
where
\begin{equation} 
S=\frac{{\rm i}}{2l_{BR}}(y\sigma_x-x\sigma_y)-\frac{{\rm i}}{2l_{D}}(x\sigma_x-y\sigma_y)
\end{equation}
is a transformation matrix and 
\begin{equation}
H_1=H_{D3}+ H_{\rm lin}^{(2)}+H_Z^{(2)}+H_{D3}^{(2)} 
\label{eq:h1} \end{equation}
is an effective spin-orbit interaction. In addition to the cubic Dresselhaus term $H_{D3}$, $H_1$ comprises the following parts:
\begin{eqnarray} 
H_{\rm lin}^{(2)}&=&\frac{\hbar^2}{4 m}\left(l_D^{-2}-l_{BR}^{-2}\right)\sigma_z (x k_y- y k_x),\label{eq:h1_1}\\
H_Z^{(2)}&=&-\mu B \sigma_z (x h_1^x+y h_1^y),\label{eq:h1_2}\\
H_{D3}^{(2)}&=&-\frac{\gamma_c}{2 l_{BR}}\left[ 4 k_x k_y -\sigma_z (\{y,k_y k_x^2\}-\{x,k_x k_y^2\})\right]\nonumber\\
&-&-\frac{\gamma_c}{2 l_D}\left[{\bf k}^2+\sigma_z (\{x,k_y k_x^2\}-\{y,k_x k_y^2\})\right].
\label{eq:h1_3}
\end{eqnarray}
Higher order terms and a constant were omitted in $H_1$. The curly brackets denote the anticommutator, while ${\bf h_1}$ is an effective spin-orbit vector specified below.

For the following discussion the symmetries of the terms in Eq.~\eqref{eq:h1} are important. First, each term has a definite time reversal symmetry: $H_Z^{(2)}$ is antisymmetric, while the other terms are time reversal symmetric. Second, the spatial symmetry of a particular term is defined by a combination of $x,y,k_x,$ and $k_y$ it contains. To exploit the spatial symmetry of the confinement, Eq.~\eqref{eq:doubledot}, we rotate the (originally crystallographic axes) coordinates such that the new $\hat{x}$ lies along ${\bf d}$. The coordinates change according to 
\begin{equation} x\to  x\cos\delta-y\sin\delta,\quad y\to  y\cos\delta+x\sin\delta,
\label{eq:rotation}
\end{equation}
and similarly for $k_x$ and $k_y$. The rotation leaves Eq.~\eqref{eq:h1_1} unchanged. In Eq.~\eqref{eq:h1_2} the effective linear spin-orbit couplings $h_1^x$ and $h_1^y$ acquire the following form:
\begin{eqnarray}
h_1^x&=&l_{BR}^{-1} \cos(\gamma-\delta) - l_D^{-1} \sin(\gamma+\delta),
\label{eq:h1x}\\
h_1^y&=&l_{BR}^{-1} \sin(\gamma-\delta) - l_D^{-1} \cos(\gamma+\delta).
\label{eq:h1y}
\end{eqnarray}
It is important that these couplings can be selectively turned to zero by orienting the static magnetic field ${\bf B}$ in a certain direction ($\gamma$) dependent on the orientation of the double dot ($\delta$). The result of the rotation in Eq.~\eqref{eq:h1_3} is not presented here; we will discuss only its relevant parts. 

We can obtain analytical results in reasonable quantitative agreement with the numerics in the lowest order degenerate perturbation theory by exploiting the symmetries of the problem. The orbital eigenfunctions of $H_0$, Eq.~\eqref{eq:h0}, in an in-plane magnetic field form a representation of $C_{2v}$ symmetry group.\cite{stano2005:PRB} There are four possible symmetry classes which transform upon inversions along (rotated axes) $\hat{x}$ and $\hat{y}$ as $1,x,xy$, and $y$, respectively. 
A relevant part of the double dot spectrum is shown in Fig.~\ref{fig:symmetry}. Several eigenstates are labeled by $\Gamma$ with indexes, where the bottom index denotes the spatial symmetry of the state (four symmetry groups), while upper indexes labels states within the symmetry group -- this  notation was introduced in Ref.~\onlinecite{stano2005:PRB}. In further the two lowest orbital states will play the most important role: ground state $\Gamma_1^{00}$ is symmetric both in x and y (often denoted as the {\it bonding} molecular orbital), and first excited orbital state $\Gamma_2^{10}$ is antisymmetric in x and symmetric in y ({\it antibonding}).

If a magnetic field is applied, each line in Fig.~\ref{fig:symmetry} splits into two by the Zeeman term lifting the degeneracy. Assuming negative g-factor and positive $B$, a spin down state (denoted by $\downarrow$) has higher energy than a spin up state ($\uparrow$). A further important consequence of a finite Zeeman energy is the anti-crossing of states $\Gamma_{1\downarrow}^{00}$ and $\Gamma_{2\uparrow}^{10}$, influence of which we take into account using the degenerate perturbation theory. Exact eigenfunctions (denoted by an overline) of the Hamiltonian $H^\prime$  can be written as a combination of the solutions of $H_0$ (denoted by $\Gamma$ as in Fig.~\ref{fig:symmetry}): the three lowest states, in the lowest order of the degenerate perturbation theory, are
\begin{eqnarray}
\overline{\Gamma}_{1\uparrow}^{00} &\approx& \Gamma_{1\uparrow}^{00}+
\frac{\langle \Gamma_{2\downarrow}^{10} H_1 \Gamma_{1\uparrow}^{00}\rangle }{E_{1\uparrow}^{00}-E_{2\downarrow}^{10}} \Gamma_{2\downarrow}^{10}+\ldots,
\label{eq:ground state}\\
\overline{\Gamma}_{1\downarrow}^{00} &\approx& \alpha\Gamma_{1\downarrow}^{00}  +\beta \Gamma_{2\uparrow}^{10}+
\frac{\langle \Gamma_{4\uparrow}^{11} H_1 \Gamma_{1\downarrow}^{00}\rangle }{E_{1\downarrow}^{00}-E_{4\uparrow}^{11}} \Gamma_{4\uparrow}^{11}+\ldots,
\label{eq:spin state}\\
\overline{\Gamma}_{2\uparrow}^{10} &\approx& \alpha\Gamma_{2\uparrow}^{10}  -\beta^* \Gamma_{1\downarrow}^{00}+
\frac{\langle \Gamma_{4\downarrow}^{11} H_1 \Gamma_{2\uparrow}^{10}\rangle }{E_{2\uparrow}^{00}-E_{4\downarrow}^{11}} \Gamma_{4\downarrow}^{11}+\ldots.
\label{eq:orbital state}
\end{eqnarray}
The dots denote the rest of an infinite sum through the eigenfunctions of $H_0$. The anti-crossing is described by coefficients 
\begin{eqnarray}
\beta&=&{\rm Arg}(c \,\delta E)\sin[\arctan(|4c/\delta E|)/2],\\
\alpha&=&\sqrt{1-|\beta|^2},
\end{eqnarray}
which depend on the energy difference  $\delta E=E_{1\downarrow}^{00}-E_{2\uparrow}^{10}$ and the coupling $c = \langle \Gamma_{2\uparrow}^{10} H_1 \Gamma_{1\downarrow}^{00} \rangle$ between the unperturbed crossing states. 

From the above expression for coupling $c$ it follows that the anti-crossing is caused by the part of $H_1$ with the spatial symmetry of $x$, which is the symmetry of $\Gamma_{2\uparrow}^{10}$. After the rotation, Eq.~\eqref{eq:rotation}, the only term with x symmetry in Eqs.~\eqref{eq:h1_1}-\eqref{eq:h1_3} is the first term in $H_Z^{(2)}$. Therefore, by orienting the magnetic field such that $h_1^x=0$, one can turn the anti-crossing into a crossing, $\beta=0$. Note that also $H_{D3}$ contains a term of $x$ symmetry -- this does not hinder to achieve $h_1^x=0$, but only slightly shifts the required position of the magnetic field.\cite{stano2006:PRL} Changing the anti-crossing into a crossing has profound consequences on the spin relaxation time, as was found in Ref.~\onlinecite{stano2006:PRL}. As we will see here, it is similarly important also for the electrically induced spin resonance. 

We finish this section with a note about other formulations of the unitary transformation Eq.~\eqref{eq:unitary transformation}. It was first used in the context of quantum dots in Ref.\onlinecite{aleiner2001:PRL}. There they neglected the cubic Dresselhaus term, but kept the correction of the third order in the spin-orbit couplings, which in our notation would be 
\begin{equation}
H^{(3)}_{\rm lin}=[S,H_{\rm lin}^{(2)}]/3.
\end{equation}
This term, which we neglected, together with Eq.\eqref{eq:h1_1}, were there interpreted as a vector potential of a spin-orbit originated magnetic field. 

If the potential is harmonic ($d=0$ in out model), the unitary transformation can be generalized to remove explicitly also the lowest order mixed Zeeman-spin-orbit term $H_Z^{(2)}$ at the expense of parameters of the Hamiltonian (like mass) becoming spin dependent. However, this possibility is specific to the potential form and nothing can be done with the cubic Dresselhaus term. 

An elegant form of the unitary transformation together with the perturbation theory is worked out in Ref.~\onlinecite{golovach2006:PRB}, where an effective Hamiltonian for a set of degenerate states is derived in a compact form using an inverse of Liouville superoperator. However the inverse is not known for any other potential than harmonic if the Zeeman term is present and is not known at all for the cubic Dresselhaus term.

The effective Hamiltonian presented here is independent of the confinement potential form and reveals the symmetry of the spin-dependent perturbations. In a symmetric potential, such as our double dot, simply by inspecting the symmetry of the terms allows to identify the term responsible for certain process (such as spin relaxation, or electrically induced transition). Formulas in Eqs.~\eqref{eq:h1_1}-\eqref{eq:h1_3} hold also if an out of plane component of the magnetic field $B_z$ is present, provided that (i) the operator ${\bf k}$ is promoted to contain also the vector potential of this component ${\bf k} \to -{\rm i}\hbar \boldsymbol{\nabla}+e {\bf A}(B_z)$ and (ii) there is an additional contribution to $H_Z^{(2)}$, proportional to $B_z$, see Ref.~\onlinecite{stano2006:PRB} for its form.

Using the effective spin-orbit Hamiltonian and the approximations for the eigenstates, we now quantify individual contributions of oscillating fields to the matrix element $\Omega_{ba}$. We will show where these contributions originate and how they can be used to control the electron spin and orbital resonance.

\section{Matrix elements: spin resonance}
\label{sec:matrix spin}
 
As the spin resonance we denote a situation, when the two resonant states are the ground state $\overline{\Gamma}_{1\uparrow}^{00}$ and its Zeeman split counterpart $\overline{\Gamma}_{1\downarrow}^{00}$. In this case we tag the matrix element by subscript ``spin'', 
\begin{equation}
\Omega_{\rm spin}^F=\langle \overline{\Gamma}_{1\uparrow}^{00} |\hat{\Omega}^F|\overline{\Gamma}_{1\downarrow}^{00}\rangle,
\label{eq:spin definition}
\end{equation}
where the upperscript $F$ will stand for the particular part of $\hat{\Omega}$ we consider. But before dealing with the specific oscillating field Hamiltonian, we remind the Van Vleck cancellation that occurs in Eq.~\eqref{eq:spin definition} due to the time reversal symmetry. Consider a general Hermitian operator $\hat{O}$. Let us write the matrix element in the following form:
\begin{equation}
\langle \overline{\Gamma}_{1\uparrow}^{00} |\hat{O}| \overline{\Gamma}_{1\downarrow}^{00}\rangle = 
\langle \Gamma_{1\uparrow}^{00} |\hat{O}| \Gamma_{1\downarrow}^{00}\rangle+
\delta O,
\end{equation}
where $\delta O$ is due to spin-orbit corrections.
If the first term is nonzero, that is the unperturbed states are coupled by $\hat{O}$, $\delta O$ can be usually neglected. If the first term vanishes, and we are away from the anti-crossing, $\beta\ll1$, the time inversion symmetry gives an important information about the matrix element $\delta O$. Indeed, if $\hat{O}$ has a definite time reversal symmetry, $T(\hat{O})=1(-1)$ when being symmetric (antisymmetric), using Eqs.~\eqref{eq:ground state}-\eqref{eq:spin state} for the matrix element in the lowest order in $H_1$ we get\cite{vleck1940:PR,khaetskii2001:PRB}
\begin{equation} \begin{split}
\delta O=&
\sum_{i,j,\sigma} \langle \Gamma_{1\uparrow}^{00} \hat{O} \Gamma^j_{i\sigma}\rangle \langle \Gamma^j_{i\sigma}  H_1 \Gamma_{1\downarrow}^{00}\rangle\times\\
 &\times \left( \frac{1}{E^{00}_{1\downarrow}-E^j_{i\sigma}}-
\frac{T(H_1)T(\hat{O})}{E^{00}_{1\uparrow}-E^j_{i,-\sigma}}\right),
\label{eq:van vleck}
\end{split} \end{equation}
where $i$ denotes the symmetry class, $j$ denotes, for brevity, both upper orbital indexes,  and $\sigma$ denotes the spin. 
In this lowest order, the contributions from the constituents of $H_1$ are additive and can be considered separately. Therefore the first order contributions of the terms with the same time reversal symmetry as $\hat{O}$ [that is if $T(H_1)T(\hat{O})=1$] will be suppressed by a factor of order of $\mu B/E_0$, compared to matrix elements such as Eq.~\eqref{eq:spin definition}, but between states with different spatial indexes. Near the anti-crossing the terms containing coefficients $\alpha$ and $\beta$ dominate other terms in Eqs.~\eqref{eq:spin state}-\eqref{eq:orbital state} and the matrix elements are then proportional to these coefficients -- the Van Vleck cancellation does not occur.

These general results can be applied to the spin resonance due to magnetic and electric fields. The oscillating magnetic field [$\hbar \hat{\Omega}=\mu  \mathcal{B}_z \sigma_z$] couples the unperturbed states: 
\begin{equation} \Omega_{\rm spin}^{\mathcal{B}_z}=\alpha \mu \mathcal{B}_z,
\label{eq:O spin B}
\end{equation}
so that we can neglect the spin-orbit contribution to the matrix element, $\delta \Omega$.

On the other hand, the electric field dipole operator ($\hbar \hat{\Omega}=e \boldsymbol{\mathcal{E}}.{\bf r}$) does not couple the unperturbed states. As $\hat{\Omega}$ is now time reversal symmetric, the contributions of all terms in $H_1$ but $H_Z^{(2)}$ are suppressed. For the electric field along the rotated $\hat{x}$ axis the matrix element at the anti-crossing is 
\begin{equation}
\Omega_{\rm spin}^{\mathcal{E}_x}=\beta e \mathcal{E}_x \overline{X}_1^\dagger.
\label{eq:O spin Exb}
\end{equation}
Away from the anti-crossing, 
\begin{equation} \begin{split}
\Omega_{\rm spin}^{\mathcal{E}_x}&=-e \mathcal{E}_x h_1^x \mu B \sum_j |\overline{X}_j|^2  \frac{2(E_2^j-E_1^{00})}{(E_2^j-E_1^{00})^2-4(\mu B)^2}.
\label{eq:O spin Ex}
\end{split} \end{equation}
The spatial symmetry (here $x$) of the dipole operator selects only eigenfunctions of symmetry $x$ in the sum. Only $H_Z^{(2)}$, Eq.~\eqref{eq:h1_2}, contains a term of x symmetry, proportional to $h_1^x$. In the above sum each state $j$ (with energy $E_2^j$) contributes proportionally to its dipole matrix element $\overline{X}_j$. To get the analytical result close to numerics one needs to include the two lowest eigenfunctions in the sum in Eq.\eqref{eq:O spin Ex}. 

If the electric field is along the rotated $\hat{y}$ axis, the anti-crossing does not influence the matrix element, since y dipole operator of the electric field does not couple the ground and anti-crossing states. Then, an analogous expression to Eq.~\eqref{eq:O spin Ex} holds at (up to factor $\alpha$ multiplying some terms in the sum) or away from the anti-crossing:
\begin{equation} \begin{split}
\Omega_{\rm spin}^{\mathcal{E}_y}&=-e \mathcal{E}_y h_1^y \mu B \sum_j |\overline{Y}_j|^2  \frac{2(E_4^j-E_1^{00})}{(E_4^j-E_1^{00})^2-4(\mu B)^2}.
\label{eq:O spin Ey}
\end{split} \end{equation}
Here it is enough to include just the lowest eigenfunction of $y$ symmetry in the sum. The dipole elements and energy differences, computed by approximating the unperturbed functions $\Gamma$ by symmetrized single dot orbitals,\cite{stano2005:PRB} are summarized in Tab.~\ref{tab:expression}.

\begin{figure}
\centerline{\psfig{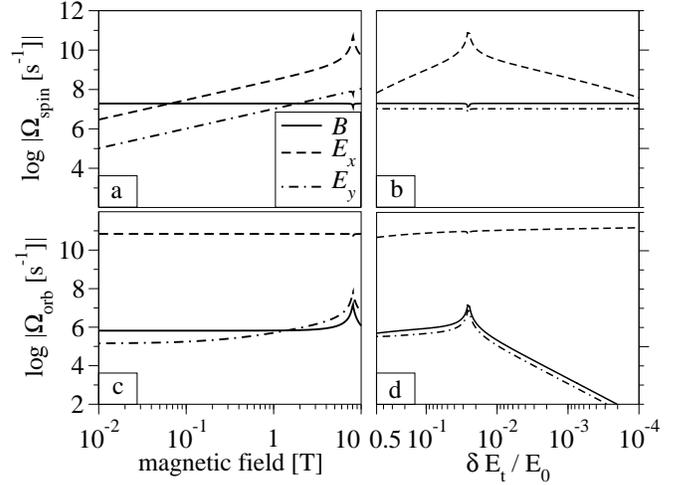}}
\caption{Calculated matrix elements between the resonant states due to magnetic and electric oscillating fields. The two upper panels, (a) and (b), show the matrix elements $\Omega_{\rm spin}$ for the spin resonance, while the two lower panels show orbital resonance elements $\Omega_{\rm orb}$. On the left, in (a) and (c) the elements are functions of the static magnetic field, with a fixed tunneling energy of 20\% of the confinement energy. On the right, in (b) and (d) the elements are functions of the tunneling energy at a fixed magnetic field $B=1$ T. The dots are oriented along [100], while the static magnetic field lies along [010].}
\label{fig:overlaps}
\end{figure}

\begin{table} \begin{tabular}{|c|c|c|c|c|c|}
\hline
&definition&unit&expression&$\mathcal{D}\ll 1$&$\mathcal{D}\gg 1$\\
\hline
$\overline{X}_1$&$\langle \Gamma_2^{10} | x | \Gamma_1^{00}\rangle$&$l_0$&$ 
\frac{\mathcal{D}}{\sqrt{1-e^{-2\mathcal{D}^2}}}$&$\frac{1}{\sqrt{2}}$&$\mathcal{D}$\\
$\overline{X}_2$&$\langle \Gamma_2^{31} | x | \Gamma_1^{00}\rangle$&$l_0$&-&$-\frac{3\mathcal{D}^2}{4}$&$\frac{1}{\sqrt{2}}$\\
$\overline{Y}_1$&$\langle \Gamma_4^{11} | y | \Gamma_1^{00}\rangle$&$l_0$&$ 
\frac{1}{\sqrt{2}}$&$ 
\frac{1}{\sqrt{2}}$&$ 
\frac{1}{\sqrt{2}}$\\
$\overline{XY}$&$\langle \Gamma_3^{21} | k_x k_y | \Gamma_1^{00}\rangle$&$l_0^{-2}$&$-\frac{\mathcal{D}}{\sqrt{2}}\frac{e^{-\mathcal{D}^2}}{\sqrt{1-e^{-2\mathcal{D}^2}}}$&
$-\frac{1}{2}$&$-\frac{\mathcal{D}}{\sqrt{2}}e^{-\mathcal{D}^2}$\\
&$E_2^{10}-E_1^{00}$&$E_0$&$2 \delta E_t$&1&$\mathcal{D} e^{-\mathcal{D}^2}$\\
&$E_2^{31}-E_1^{00}$&$E_0$&-&3&1\\
&$E_4^{11}-E_1^{00}$&$E_0$&1&1&1\\
\hline
\end{tabular}
\caption{Analytical approximations for the dipole matrix elements and energy differences. For each quantity the definition, unit, expression, and limits for small and large interdot distances are given. In some cases the expression is too lengthy and only the asymptotics are given. The expression for $\delta E_t$ is given in Ref.~\onlinecite{stano2005:PRB}. The interdot distance measured in the units of the confinement length is used, $\mathcal{D}=d/l_0.$}
\label{tab:expression}
\end{table}

Fully numerical results for the matrix elements as a function of the magnetic field are shown in Fig.~\ref{fig:overlaps}a. The matrix element of the magnetic field is constant, up to a narrow region of suppression due to $\alpha$, since it depends only on the strength of the oscillating magnetic field, Eq.~\eqref{eq:O spin B}. The matrix elements of the electric field [Eqs.~\eqref{eq:O spin Ex} and \eqref{eq:O spin Ey}] are proportional to the Zeeman energy $\mu B$ --  the spin resonance is more sensitive to electrical disturbance as the magnetic field grows, while at zero magnetic field the electric field is ineffective. At the anti-crossing, $\Omega_{\rm spin}^{\mathcal{E}_x}$ is strongly enhanced (by two orders of magnitude) and described by Eq.~\eqref{eq:O spin Exb}, while $\Omega_{\rm spin}^{\mathcal{E}_y}$ develops a small dip similar to $\Omega_{\rm spin}^{\mathcal{B}_z}$.

It can be seen in Fig.~\ref{fig:overlaps}b, where the matrix elements are functions of the tunneling energy, that the spin resonance is much more sensitive to the electric field along the double dots $x$ axis than to a perpendicular field. This difference is strengthened at the anti-crossing. Only in the truly single dot case ($d=0$ or $d=\infty$), the electric field influence is isotropic. We can also conclude from the single dot values that the matrix elements of magnetic field of 1 mT and electric field of $10^3$ V/m are comparable in magnitude in the static magnetic field of order of Tesla. This means that in the experiment,\cite{koppens2006:N} where no electrically induced signal was observed, the electric field is likely considerably lower than the estimated value of $10^4$ V/m. 

\begin{figure}
\centerline{\psfig{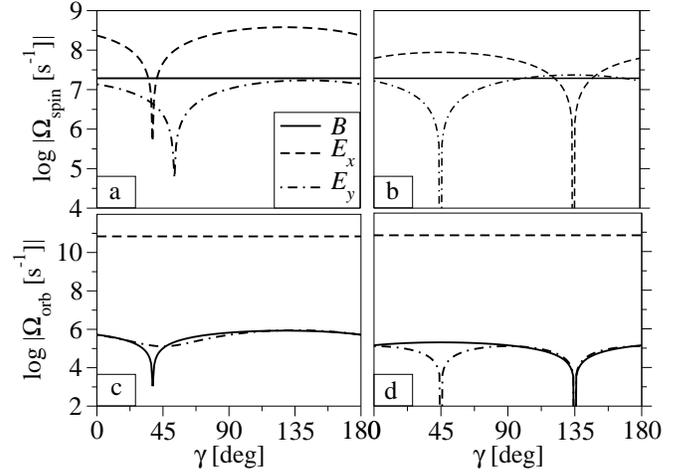}}
\caption{Calculated matrix elements for the spin [upper two panels (a) and (b)] and the orbital [lower two panels (c) and (d)] resonance due to oscillating magnetic and electric fields as functions of $\gamma$, the orientation of the static magnetic field, $B=1$ T. The tunneling energy is 20\% of the confinement energy. On the left, in (a) and (c) the dots are oriented along [100], that is $\delta=0$. On the right, in (b) and (d) the dots are oriented along [110], $\delta=45^\circ$.}
\label{fig:angle}
\end{figure}

Similarly to the spin relaxation rates,\cite{stano2006:PRL,olendski2006:CM} the matrix element of the resonant electric field is highly anisotropic. The possible control over the resonance is demonstrated in Fig.~\ref{fig:angle}a, where the matrix elements are shown as functions of the orientation of the static magnetic field. The magnetic field matrix element is independent on $\gamma$, as follows from Eq.~\eqref{eq:O spin B}. The electric field matrix elements are anisotropic, with the dependence given by the effective spin-orbit couplings $h_1^x$ and $h_1^y$. By proper orientation of the static magnetic field it is thus possible to turn off the contribution due to the electric field pointed along a given direction. In particular, the electric field along $\hat{x}$ is not effective ($h_1^x=0$) at $\gamma=\arctan(l_D/l_{BR})\approx 38^\circ$. The electric field along $\hat{y}$ is ineffective if $\gamma=\arctan(l_{BR}/l_D)\approx 58^\circ$, since here $h_1^y=0$. These conditions were obtained from Eqs.~\eqref{eq:h1x} and \eqref{eq:h1y} by putting $\delta=0$ (the dots oriented along [100]).  Different orientation of the dots changes the conditions for the effective spin-orbit couplings to be zero. For example, in Fig.~\ref{fig:angle}b, the dots are oriented along [110], that is $\delta=45^\circ$ and the effective couplings $h_1^x$ and $h_1^y$ are zero at $\gamma=45^\circ$ and $135^\circ$, respectively, independent on the spin-orbit parameters. If the electric field points along a general direction,  it is still possible to turn off the matrix element by properly orienting the magnetic field. However, in a general case the desired position of the magnetic field is defined not only by the effective couplings $h_1^x$ and  $h_1^y$, but by all terms in Eqs.~\eqref{eq:O spin Ex}-\eqref{eq:O spin Ey}. 

In the easy passage configuration, defined by $h_1^x=0$, the spin relaxation time does not suffer a drastic suppression due to the anti-crossing, as was shown in Ref.~\onlinecite{stano2006:PRL} We see that in addition to that here the spin resonance is insensitive to otherwise most effective electric field component -- along $\hat{x}$. Such electric fields are inevitably present if the spin qubit is manipulated by an on-chip generated magnetic field.\cite{koppens2006:N} On the other hand, on-chip manipulations seem inevitable in a scalable system, where it must be possible to address the qubits selectively. The easy passage configuration thus protects the spin against the electric field and provides a stable Rabi frequency over a wide range of parameters values, if the qubit is manipulated by an oscillating magnetic field.

\section{Matrix elements: orbital resonance.}
\label{sec:matrix orbital} 

As the orbital resonance we consider the case when the resonant states are the two lowest orbital states and the matrix element is 
\begin{equation}
\Omega_{\rm orb}^F=\langle \overline{\Gamma}_{1\uparrow}^{00} |\hat{\Omega}^F|\overline{\Gamma}_{2\uparrow}^{00}\rangle.
\label{eq:orbital definition}
\end{equation}
A similar suppression as in Eq.~\eqref{eq:van vleck} takes place also now, if the operator $\hat{O}$ acts only in the spin subspace (that is, it is the Zeeman term). This suppression again favors the contribution due to $H_Z^{(2)}$ compared to the rest of $H_1$. If the anti-crossing dominates, the matrix element due to $\mathcal{B}_z$ is $\Omega_{\rm orb}^{\mathcal{B}_z}=-\beta^\dagger\mu \mathcal{B}_z$, while away from the anti-crossing
\begin{equation} 
\Omega_{\rm orb}^{\mathcal{B}_z}= -\mu \mathcal{B}_z  h_1^x \overline{X}_1 \frac{(E^{10}_2-E_1^{00})^2}{(E^{10}_2-E_1^{00})^2-4(\mu B)^2}.\\
\label{eq:O orb B}
\end{equation}
Contrary to the case of electrically induced spin resonance, the oscillating magnetic field can induce transitions also at zero static magnetic field, as seen in Fig.~\ref{fig:overlaps}c. 
However, the matrix element of the magnetic field is overlaid if the electric field in the x direction is present, since such electric field is much more efficient for the orbital resonance,
\begin{equation} 
\Omega_{\rm orb}^{\mathcal{E}_x}=e \mathcal{E}_x \overline{X}_1,
\end{equation}
because it couples unperturbed states directly.

If the electric field is oriented along $\hat{y}$, it is much less effective, because the linear spin-orbit terms do not contribute in the first order. Here, for a non-zero matrix element in Eq.\eqref{eq:orbital definition}, the perturbation $H_1$ has to contain a term which is spin diagonal with spatial symmetry $xy$. The only such in $H_1$ is the term originating in the first term of $H_{D3}^{(2)}$, Eq.~\eqref{eq:h1_3}. After the rotation of the coordinate system this term is $ -(2\gamma_c/l_{BR})\cos(2\delta) k_x k_y$, leading to the matrix element 
\begin{equation} \begin{split}
\Omega_{\rm orb}^{\mathcal{E}_y}&=e \mathcal{E}_y \overline{Y}_1 \frac{\gamma_c}{l_{BR}}\cos2\delta \overline{XY} \frac{4(E_1^{00}-E_2^{10})}{(E_4^{11}-E_1^{00})^2-(E_2^{10}-E_1^{00})^2}.
\label{eq:O orb Ey}
\end{split} \end{equation}
In small magnetic fields ($\lesssim 1$ T) this contribution dominates the matrix element compared to the contributions from other parts in $H_1$, such as $H_Z^{(2)}$, contributing in the second order. Note that there is no term with appropriate symmetry (spin diagonal, spatially $xy$) of $H_1$ coming from a mixture of $H_D$ and $H_{D3}$, making $\Omega_{\rm orb}^{\mathcal{E}_y}$ a specific effect due to the mixed cubic Dresselhaus and Bychkov-Rashba interactions. This example demonstrates the usefulness of information about the symmetry contained in Eqs.~\eqref{eq:h1_1}-\eqref{eq:h1_3}. By simple inspection of the symmetry we can tell immediately which term needs to be considered for a specific situation.

The dependence of the matrix elements on the static magnetic field orientation $\gamma$ is shown in Fig.~\ref{fig:angle} c and d. The magnetic field matrix element is proportional to $h_1^x$, see Eq.~\eqref{eq:O orb B}. The direct coupling through the electric field along $\hat{x}$ is independent on $\gamma$. The matrix element of the electric field along $\hat{y}$, as given in Eq.~\eqref{eq:O orb Ey}, is independent on $\gamma$ and can not be put to zero by changing the magnetic field orientation -- as seen in Fig.~\ref{fig:angle}c. However, there is some dependence to be seen and the dependence is striking for a different dots' orientation. The reason is that Eq.~\eqref{eq:O orb Ey} is the dominant contribution to the matrix element only up to a certain value of the static magnetic field -- in higher fields the second order contribution from $H_Z^{(2)}$ will dominate. Since there is already a visible dependence in Fig.~\ref{fig:angle}c, we can estimate the crossover magnetic field to be 1 Tesla, for our parameters. In Fig.~\ref{fig:angle}d, the contribution of Eq.~\eqref{eq:O orb Ey} is zero exactly, since $\delta=45^\circ$. Therefore the second order contribution to the matrix element coming from $H_Z^{(2)}$ is seen. The possible dependence of the matrix element on $\gamma$ can decide whether the matrix element is induced by linear spin-orbit terms (depends on $\gamma$), or the mixed cubic-linear terms (does not depend on $\gamma$). This could be used as a detection for the presence of the cubic Dresselhaus term. Unless the electric field is positioned exactly along y axis, no oscillating magnetic field influence or anisotropy can be observed due to high effectiveness of the electric field along $\hat{x}$. 

After having analyzed means of control over the field matrix element, or, in another words, Rabi frequency, we will now study the steady state solution of the density matrix. We will show that the Rabi frequency and decoherence, which have been obtained in Refs.~\onlinecite{koppens2006:N,petta2005:S} from the observation of the decaying Rabi oscillations, can be obtained alternatively from the steady state current measurement.

\section{Resonant field influence in the steady state}

In this Section we are interested in the steady state solution of the density matrix, denoted by $\overline{\rho}$ and defined as the solution with constant occupations
\begin{equation}
(\partial^{\rm ph}_t+\partial^{\rm of}_t)\overline{\rho}_{ii}=0,\quad \forall i,
\label{eq:steady state dm}
\end{equation}
where the two contribution to the time derivative are those in Eqs.~\eqref{eq:phonon channel} and \eqref{eq:field channel}. Even though it is not currently measurable in a single electron system, we include in our list of interesting steady state parameters the absorption,
\begin{equation}
W=\partial^{\rm of} \sum_i E_i \overline{\rho}_{ii},
\end{equation}
defined as the energy gain of the electron due to the oscillating field.

After the decay of the Rabi oscillations, the system is in the steady state, where the occupations are constant. In this case the time derivative of the density matrix due to the oscillating field Eqs.~\eqref{eq:field channel} can be simplified to (see Ref.~\onlinecite{fabian2007:APS} for the derivation)
\begin{equation}
\label{eq:induced channel}
\partial_t^{\rm of}\, \rho_{aa}=-\partial_t^{\rm of}\, \rho_{bb}=2(\rho_{bb}-\rho_{aa}) J,
\end{equation}
where the induced rate
\begin{equation}
J=\frac{|\Omega_{ba}|^2}{4}\frac{\gamma_{ba}}{\Delta^2+\gamma_{ba}^2}.
\label{eq:induced rate}
\end{equation}
The zero time derivative of the occupations in the steady state  can be interpreted as a balance between two competing processes -- relaxation [Eqs.\eqref{eq:phonon channel}] which drives the system towards the thermodynamical equilibrium ($\rho_{bb}/\rho_{aa}=\Gamma_{ab}/\Gamma_{ba}$) and oscillating field induced transition [Eq.\eqref{eq:induced channel}] equilibrating occupations of the resonant states ($\rho_{bb}=\rho_{aa}$). The effectiveness of the oscillating field in driving the system out of thermal equilibrium is characterized by the induced rate $J$, Eq.~\eqref{eq:induced rate}. Going away from resonance the oscillating field is less effective in influencing the system, reflected by the (Lorentzian shape) decay of the induced rate.

Our numerical strategy to obtain the steady state density matrix $\overline{\rho}$ is as follows: We diagonalize the coupled dots electron Hamiltonian, Eq.~\eqref{eq:hamiltonian},\cite{stano2005:PRB} and compute the relaxation rates using Fermi's Golden rule.\cite{stano2006:PRB} We choose a pair of resonant states, \{a, b\}, and after evaluating $\Omega_{ba}$ we find the induced rate according to Eq.~\eqref{eq:induced rate}. Finally, we find the steady state density matrix by solving the set of linear equations defined by Eq.~\eqref{eq:steady state dm}. A different method, with the oscillating field treated exactly, was used for single dot in intense oscillating fields,\cite{jiang2006:JAP} three orders of magnitude larger than the fields considered here.

We can analytically reproduce the numerical results by the two state approximation discussed in the above. The physics is then characterized by the number 
\begin{equation}
J^r_0=\Gamma_{ba}^{-1} J|_{\omega=\omega_{ba}}=|\Omega_{ba}|^2\frac{1}{4\gamma_{ba}\Gamma_{ba}},
\end{equation}
which is the induced rate at the resonance, measured in the units of the relaxation rate between the resonant states.

Two limits can be identified, according to $J^r_0$. If the induced rate dominates the relaxation, $J^r_0\gg 1$, the occupations of the two resonant states are close to being equal, while if $J_0^r \ll 1$, the system is close to the thermal equilibrium.
The interpretation of $2J$ as the electron outscattering rate due to the oscillating field, as it follows from Eqs.~\eqref{eq:induced channel}, is reassured by the result form the absorption. We expect the absorption to be proportional to a transition rate from the excited state to the ground state times the energy dissipated at this transition. If $J_0^r \ll 1$ the transition rate is $2J$.  In the opposite limit, $J_0^r \gg 1$, the outscattering due to the oscillating field is strong and the transition rate for the dissipation is limited by the relaxation rate. The frequency full widths at half maximum (FWHM) also differ for the two limits -- see Tab.~\ref{tab:results} for analytical results.

\begin{table}
\begin{tabular}{|c|c|c|c|}
\hline
a&steady state&at resonance&FWHM ($\delta\omega_{1/2}^2$)\\
\hline
$\overline{\rho}_{bb}$&$\frac{J+\tau \Gamma_{ba}}{2J+\Gamma_{ba}(1+\tau)}$&$\frac{J_0^r+\tau}{2J_0^r+1+\tau}$&
$\frac{8J_0^r(1+J_0^r)+4\tau(1+\tau+3J_0^r)}{J_0^r-\tau(1+\tau+3J_0^r)}\gamma_{ba}^2$\\
$J$&$\frac{|\Omega_{ba}|^2\gamma_{ba}}{4\Delta^2+4 \gamma_{ba}^2}$&$|\Omega_{ba}|^2/4\gamma_{ba}$&$4\gamma_{ba}^2$\\
$W$&$E_{ba} J \frac{2(1-\tau)}{1+\tau+2 J/\Gamma_{ba}}$&
$E_{ba} J \frac{2(1-\tau)}{1+\tau+2 J_0^r}$&
$\frac{4(1+\tau+2J_0^r)}{1+\tau}\gamma_{ba}^2$\\
\hline
\end{tabular}
\begin{tabular}{|c|c|c|c|}
\hline
b&limit&at resonance&FWHM ($\delta\omega_{1/2}^2$)\\
\hline
$\overline{\rho}_{bb}$&$J_0^r\gg 1$&$1/2-(1-\tau)/2J_0^r$&$2 |\Omega_{ba}|^2 \gamma_{ba}/\Gamma_{ba}(1-3\tau)$\\
$\overline{\rho}_{bb}$&$J_0^r\ll 1$&$\frac{\tau}{1+\tau}+J_0^r(1-\tau)/(1+\tau)^2$&4$\gamma_{ba}^2$\\
$W$&$J_0^r\gg 1$&$E_{ba} \Gamma_{ba} (1-\tau)$&
$2 |\Omega_{ba}|^2 \gamma_{ba}/\Gamma_{ba}(1+\tau)$\\
$W$&$J_0^r\ll 1$&$2 E_{ba} J^r(1-\tau)/(1+\tau)$&
4$\gamma_{ba}^2$\\
\hline
\end{tabular}
\caption{(a) Steady state, value at resonance, and frequency full width at half maximum (FWHM) $\delta\omega_{1/2}$ squared for the excited state occupation $\overline{\rho}_{bb}$, the induced rate $J$, and absorption $W$. Note that the FWHM of the excited population is defined only if the temperature is low enough such that $J_0^r\geq \tau(1+\tau)/(1-3\tau)$. (b) The value at the resonance, and frequency full width at half maximum of the excited population and absorption in the two limits.}
\label{tab:results}
\end{table}

\begin{figure}
\centerline{\psfig{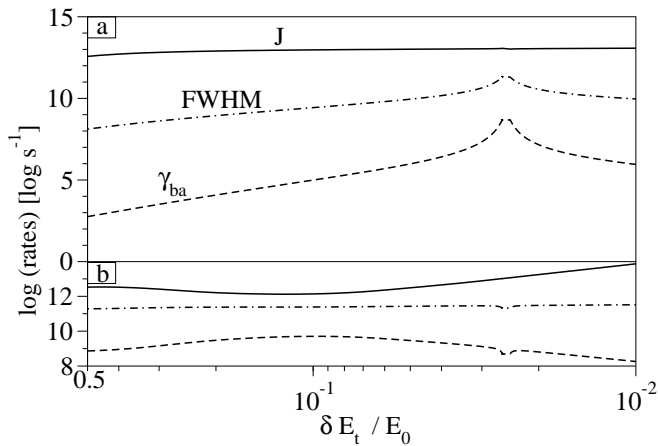}}
\caption{Calculated induced rate $J$ at resonance (solid), decoherence $\gamma_{ba}$ (dashed), and the FWHM of the excited population (dot-dashed) as functions of the ratio of the tunneling energy $\delta E_t$ and the confinement energy $E_0$ for (a) spin resonance and (b) orbital resonance. The static in-plane magnetic field is $B=1$ T.  If the solid line is above (under) the dashed one, it means that $J_0^r > 1$ ($J_0^r < 1$). The dots are oriented along [100], while the static magnetic field lies along [010].}
\label{fig:current}
\end{figure}

Figure \ref{fig:current} presents our numerical results for induced rate, excited population width, and decoherence as functions of the tunneling energy for the spin and orbital resonance. Both resonances are in the regime of $J_0^r\gg 1$, where the decoherence is revealed by the FWHM of the induced rate, see Tab.~\ref{tab:results}, while the relaxation rate can be obtained if both the induced rate at resonance and FWHM of the excited population are known, too.
 Due to Eq.~\eqref{eq:decoherence}, the relaxation rate is indiscernible from the decoherence in the figure and $J^r_0$ can be directly determined. For the spin resonance $J^r_0$ varies between $10^5$ and $10^{11}$ -- the limit expressions in Tab.~\ref{tab:results} are then exact with this precision. The upward dips in FWHM and the decoherence rate are due to the anti-crossing of the spin and orbital states.\cite{stano2005:PRB} It is interesting that the induced rate is not influenced by the anti-crossing. This is because both the square of the matrix element and the decoherence (equal to the relaxation) in Eq.~\eqref{eq:induced rate} depend on the anti-crossing in the same way and the contributions cancel. Also note that the rates characterizing the oscillating field are very different in the transient and the steady state regime. While the steady state characteristic rate $J$ is $\sim10^{13}$ s$^{-1}$, looking at Fig.~\ref{fig:overlaps}b one can see that the Rabi frequency for the same parameters is only $\sim 10^8$ s$^{-1}$.

Compared to the spin resonance, the orbital resonance is much less sensitive to the anti-crossing, since only in a very narrow region at the anti-crossing the relaxation rate acquires a factor of one half.\cite{stano2006:PRB} One also sees that $J^r_0$ is smaller, meaning we are closer to the regime of $J^r_0 <1$ which can be reached by lowering the amplitude of the oscillating electric field. In that regime, the decoherence can be obtained from the FWHM of the excited population or from the induced rate.

We finish this section by summarizing that after identifying the appropriate regime of high or low induced rate one can obtain the decoherence and Rabi frequency using expressions from Tab.\ref{tab:results} provided one can measure the induced rate $J$ and the excited state population $\overline{\rho}_{bb}$ (and their full widths). In turn, these two can be measured if the dot is connected to leads and the current flows through the dot, as shown theoretically in Ref.~\onlinecite{engel2001:PRL}. In Ref.~\onlinecite{fabian2007:APS} it is shown even on a simpler model that the measurement can be done by changing the coupling between the dot and the leads. Namely, for small coupling the current is proportional to the excited state population, while for large coupling the current measures the induced rate.

\section{Conclusions}

We have studied electrically and magnetically induced spin and orbital resonance of a single electron confined in coupled lateral quantum dots. We have taken into account the relaxation and decoherence due to an acoustic phonon environment, with the rates computed by Fermi's Golden rule. Resonant oscillating electromagnetic fields are capable to induce transitions between electron eigenstates. We have focused on the oscillating field matrix elements, equal to the Rabi frequency, for spin and orbital resonance. 

We have given an effective spin-orbit Hamiltonian which allows to quantify the spin-orbit influence on the matrix element using symmetry considerations. Specifically, for electrically induced spin resonance, we have shown how the spin-orbit anisotropy allows to control the matrix element by the strength and orientation of the static magnetic field. These conclusions give hints for optimal quantum dot configurations for the case when: (i) the spin is manipulated by an oscillating electric field, whereas its influence is desired to be maximized and (ii) the spin manipulated by an oscillating magnetic field, when the effect of the electric field on the spin is desired to be minimized. Connecting with our previous work, we have found that the easy passage provides not only long spin relaxation time, but also stability against electric field disturbances, making it a suitable arrangement for spin qubit realization.
 
In a double dot, the electric field is most effective in spin manipulation if it lies along the dots' axis and the matrix element is strongly influenced by the anti-crossing (a crossing of different spin states lifted by spin-orbit interactions). An important feature is that the electric field is less effective if the magnitude of the static magnetic field is lowered. Oscillating electric field of order of 1000 V/m can easily be more effective than oscillating magnetic field of 1 mT if the static magnetic field is of order of Tesla. For these parameters in a GaAs quantum dot the Rabi frequency of 1 GHz is achievable for the spin manipulation using an electric field.
  
In the last part we studied the influence of the resonant fields in the steady state. We proposed the induced rate as a single characteristic parameter. We have analyzed steady state occupations, induced rate, and absorption and their full widths for both spin and orbital resonance and used those results to show how to obtain decoherence and Rabi frequency from these steady state characteristics. In turn, these characteristics can be obtained from a steady state current measurement.

\acknowledgments
This work was supported by the US ONR.\\

\bibliography{../../../references/quantum_dot}

\appendix

\end{document}